\documentclass[prl,letterpaper,twocolumn,showpacs,superscriptaddress,amsmath,amsfonts,amssymb,preprintnumbers]{revtex4}
\pdfoutput=1
\usepackage[T1]{fontenc}\usepackage[latin1]{inputenc}
\usepackage{dcolumn,graphicx,color,hvfloat,booktabs,microtype,afterpage} %\graphicspath{{Figures/}}
\usepackage[charter]{mathdesign}
\renewcommand{\figurename}{Fig.}
\renewcommand{\tablename}{Table}
\makeatletter\renewcommand{\fnum@figure}[1]{\figurename~\thefigure~(color online).}\makeatother
\makeatletter\renewcommand{\fnum@table}[1]{\tablename~\thetable.}\makeatother

\usepackage[colorlinks,plainpages=false,linkcolor=blue,urlcolor=blue,citecolor=black,pdfpagemode=UseNone,pdfstartview=FitBH]{hyperref}

\newcommand{\bkfa }{\mbox{Ba$_{1-x}$K$_x$Fe$_2$As$_2$}}
\newcommand{\bfa }{\mbox{BaFe$_2$As$_2$}}
\newcommand{\loffa }{\mbox{LaFeAsO$_{1-x}$F$_x$}}
\newcommand{\soffa }{\mbox{SmFeAsO$_{1-x}$F$_x$}}
\newcommand{\coffa }{\mbox{CeFeAsO$_{1-x}$F$_x$}}
\newcommand{\gtfao }{\mbox{Gd$_{1-x}$Th$_x$FeAsO}}
\newcommand{\muSR }{\mbox{$\mu$SR}}
\newcommand{\TSDW }{\mbox{$T_{\rm m}$}}

\newcommand{\mFe }{\mbox{$m_{\rm Fe}$}}
\newcommand{\zff }{\mbox{$\nu_{\rm ZF}$}}

\newcommand{\ybcosixfour }{\mbox{YBa$_2$Cu$_3$O$_{6.45}$}}

\begin{document}

%\preprint{\textit{Preprint: \today. For internal use only, do not distribute.}}

\title{Electronic phase separation in the slightly underdoped iron pnictide superconductor \bkfa}

\author{J.\,T.~Park}\author{D.\,S.\,Inosov}
\affiliation{Max-Planck-Institut für Festkörperforschung, Heisenbergstraße 1, 70569 Stuttgart, Germany.}

\author{Ch.\,Niedermayer}
\affiliation{Laboratory for Neutron Scattering, ETHZ \& PSI, CH-5232 Villigen PSI, Switzerland.}

\author{G.\,L.\,Sun}\author{D.\,Haug}
\affiliation{Max-Planck-Institut für Festkörperforschung, Heisenbergstraße 1, 70569 Stuttgart, Germany.}

\author{N.\,B.\,Christensen}
\affiliation{Laboratory for Neutron Scattering, ETHZ \& PSI, CH-5232 Villigen PSI, Switzerland.}
\affiliation{Materials Research Department, Ris{\o} National Laboratory for Sustainable Energy, Technical University of Denmark, DK-4000 Roskilde, Denmark.}
\affiliation{Nano-Science Center, Niels Bohr Institute, University of Copenhagen, DK-2100 Copenhagen, Denmark.}

\author{R.\,Dinnebier}\author{A.\,V.~Boris}
\affiliation{Max-Planck-Institut für Festkörperforschung, Heisenbergstraße 1, 70569 Stuttgart, Germany.}

\author{A.\,J.\,Drew}
\affiliation{Department of Physics and Fribourg Center for Nanomaterials, University of Fribourg, Chemin du Mus\'{e}e 3, CH-1700 Fribourg, Switzerland.}
\affiliation{Physics Department, Queen Mary, University of London, London, E1 4NS, United Kingdom.}

\author{L.\,Schulz}
\affiliation{Department of Physics and Fribourg Center for Nanomaterials, University of Fribourg, Chemin du Mus\'{e}e 3, CH-1700 Fribourg, Switzerland.}

\author{T.~Shapoval}\author{U.~Wolff}\author{V.~Neu}
\affiliation{IFW Dresden, Institute for Metallic Materials, P.\,O.\,Box 270116, D-01171 Dresden, Germany.}

\author{Xiaoping Yang}\author{C.\,T.~Lin}\author{B.\,Keimer}\author{V.~Hinkov{\large\hyperref[CorrAuthor]{*}}}
\affiliation{Max-Planck-Institut für Festkörperforschung, Heisenbergstraße 1, 70569 Stuttgart, Germany.}

\begin{abstract}
Here we present a combined study of the slightly underdoped novel pnictide superconductor \bkfa\ by means of X-ray powder diffraction, neutron scattering, muon spin rotation (\muSR), and magnetic force microscopy (MFM). Commensurate static magnetic order sets in below $T_{\rm m}\approx70$\,K as inferred from the emergence of the magnetic (1\,0\,--3)$_{\rm O}$ reflection in the neutron scattering data and from the observation of damped oscillations in the zero-field-\muSR\ asymmetry. Transverse-field \muSR\ below $T_{\rm c}$ shows a coexistence of magnetically ordered and non-magnetic states, which is also confirmed by MFM imaging. We explain such coexistence by electronic phase separation into antiferromagnetic and superconducting/normal state regions on a lateral scale of several tens of nanometers. Our findings indicate that such mesoscopic phase separation can be considered an intrinsic property of some iron pnictide superconductors.
\end{abstract}

\keywords{superconducting materials, muon spin rotation, elastic neutron scattering, magnetic force microscopy}
\pacs{\vspace{-0.2em}74.70.-b 76.75.+i 25.40.Dn 68.37.Rt\vspace{-0.5em}}

\maketitle
Since the discovery of superconductivity (SC) with a critical temperature of $T_{\rm c}$ = 26\,K in \loffa\ \cite{Kamihara08}, layered iron pnictide superconductors have attracted much attention. Compounds exhibiting a higher $T_{\rm c}$ have been successfully synthesized, like the double-layer ``122''-compound \bkfa\ (BKFA), $T_{\rm c}=38$~K \cite{RotterPRL08}, and the single-layer ``1111''-compound \gtfao, $T_{\rm c}=56$~K \cite{Wang08}. It has been established that their parent compounds ($x=0$) are poor metals that undergo a spin-density-wave (SDW) transition below typical temperatures \TSDW\ in the range between 140 and 200\,K as seen by neutron scattering \cite{Cruz08, Zhao08, Chen08a, Su08} and local-probe methods like \muSR\ \cite{Aczel08, Jesche08, Klauss08, Goko08} and $^{57}$Fe M{\"o}ssbauer spectroscopy \cite{Klauss08, RotterPRB08}.

Despite the quick development of the field, many important physical issues are still discussed controversially, such as the ground state of the parent compound \cite{GroundState}, the pairing symmetry in the SC state \cite{PairingSymmetry}, dramatically different magnetic phase diagrams \cite{Luetkens08, Zhao08, Drew08, Chen08a, Goko08}, and so forth. In addition, it has been reported that SC and antiferromagnetic (AF) order are either well separated or coexist in the underdoped region of the phase diagram. More specifically, in \loffa, the transition between the SDW and SC was reported to be first-order-like with no coexistence between the two phases \cite{Luetkens08}, in \coffa\ the transition is more second-order-like, but still the AF and SC domes do not overlap in the phase diagram \cite{Zhao08}, whereas the coexistence of the two phases was reported in a narrow doping range in \soffa\ \cite{Drew08, DrewNiedermayer08}, and in a broader range in \bkfa\ \cite{Chen08a,Goko08}. However, there is no consensus yet about the nature of such coexistence\,---\,whether it is purely electronic or related to chemical homogeneity of the sample, and what is the characteristic spatial scale of the corresponding phases. One of the reasons for such uncertainty is that many of the samples employed so far were powders.

\begin{figure}[b]\vspace{-1em}
\includegraphics[width=\columnwidth]{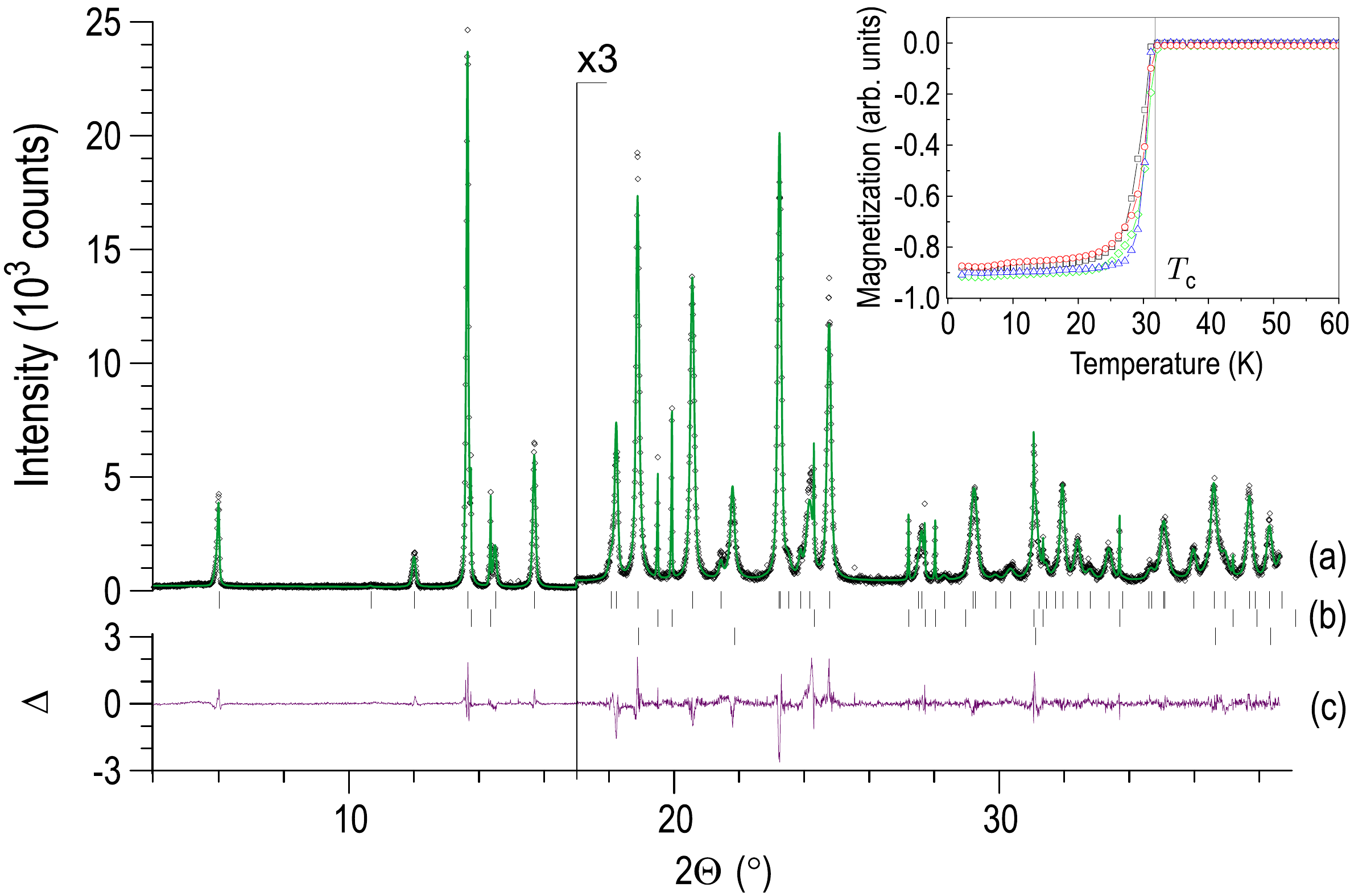}\vspace{-0.5em}
\caption{Room-temperature XRPD data. (a)~Scattered X-ray intensity at $T=300$\,K as a function of diffraction angle $2\Theta$ ($\lambda=0.7$\,\AA) fitted to the tetragonal \textit{I4/mmm} space group. For $2\Theta>17^\circ$ the plot is enlarged by a factor of three. The fit includes a few wt.\,\% of tetragonal $\beta$-tin from the flux as an impurity phase and some reflections of the brass sample holder as indicated by the reflection markers in (b). (c)~The difference $\Delta$ between the ex\-pe\-rimental points and the fitting curve. The inset shows a sharp and reproducible SC transition measured by dc susceptibility on four randomly selected samples.\vspace{-1.3em}}
\label{fig:xray}
\end{figure}

Besides, complementary experiments on nominally the same composition were often performed on samples prepared by different labs. Thus, to overcome this uncertainty we have performed dc susceptibility, X-ray powder diffraction (XRPD), neutron scattering, and \muSR\ measurements on the same BKFA samples. From these measurements, the phase separated coexistence between static magnetic order and non-magnetic (either superconducting below $T_{\rm c}$ or normal-state above $T_{\rm c}$) regions was observed, and we suggest that this phase separation is an intrinsic property of underdoped BKFA, resulting from the electronic instability to the formation of static AF islands surrounded by non-magnetic regions below the SDW transition temperature. In this respect, the iron pnictide superconductors resemble transition metal oxides, where electronic phase separation phenomena are common at various spatial scales \cite{SeparationReviews}, such as cuprates \cite{SeparationCuprates, SeparationMacro} or manganites \cite{SeparationManganates}. On the other hand, as we will subsequently show, a quantitative comparison reveals essential differences between the characteristic scales of the inhomogeneities, which might affect the macroscopic physical properties of these materials.

The single crystals of BKFA were grown using Sn as flux in a zirconia crucible sealed in a quartz ampoule filled with Ar. A mixture of Ba, K, Fe, As, and Sn in a wt. ratio of BKFA:Sn = 1:85 was heated in a box furnace up to 850$^\circ$C and kept constant for 2\,--\,4 hours to soak the sample in a homogeneous melt. The cooling rate of 3$^\circ$C/h was then applied to decrease the temperature to 550$^\circ$C, and the grown crystals were then decanted from the flux \cite{Lin08}.

Sample characterization by resistivity (not shown) and dc susceptibility measurements (see inset in Fig.\,\ref{fig:xray}) revealed a sharp SC transition at $T_{\rm c,\,onset}=(32\pm1)$~K, reproducible among different samples from the same batch. XRPD data confirm that our crystals consist of a single phase fitted well by a tetragonal \textit{I4/mmm} space group symmetry both at room temperature (see Fig.\,\ref{fig:xray}) and at $T=16$~K. Nevertheless, throughout this letter we will use the orthorhombic notation, inherited from the parent compound. The room-temperature lattice parameters of the sample, as determined from XRPD by Rietveld refinement using the fundamental parameters approach of TOPAS \cite{ChearyCoelho05}, are $a=b=3.9111(1)$\,\AA\ and $c=13.3392(6)$\,\AA. Our density functional calculations, using the projected-augmented-wave method in the framework of the generalized gradient approximation \cite{DFT}, have confirmed that the width of the diffraction peaks is comparable with that expected for homogeneous statistical distribution of the dopant atoms. From the functional dependency of the lattice parameters on doping \cite{RotterPangerl08}, the average potassium content of $x=0.41$ could be determined, in agreement with the results of our energy dispersive X-ray analysis.

\begin{figure}[t]
\includegraphics[width=1.02\columnwidth]{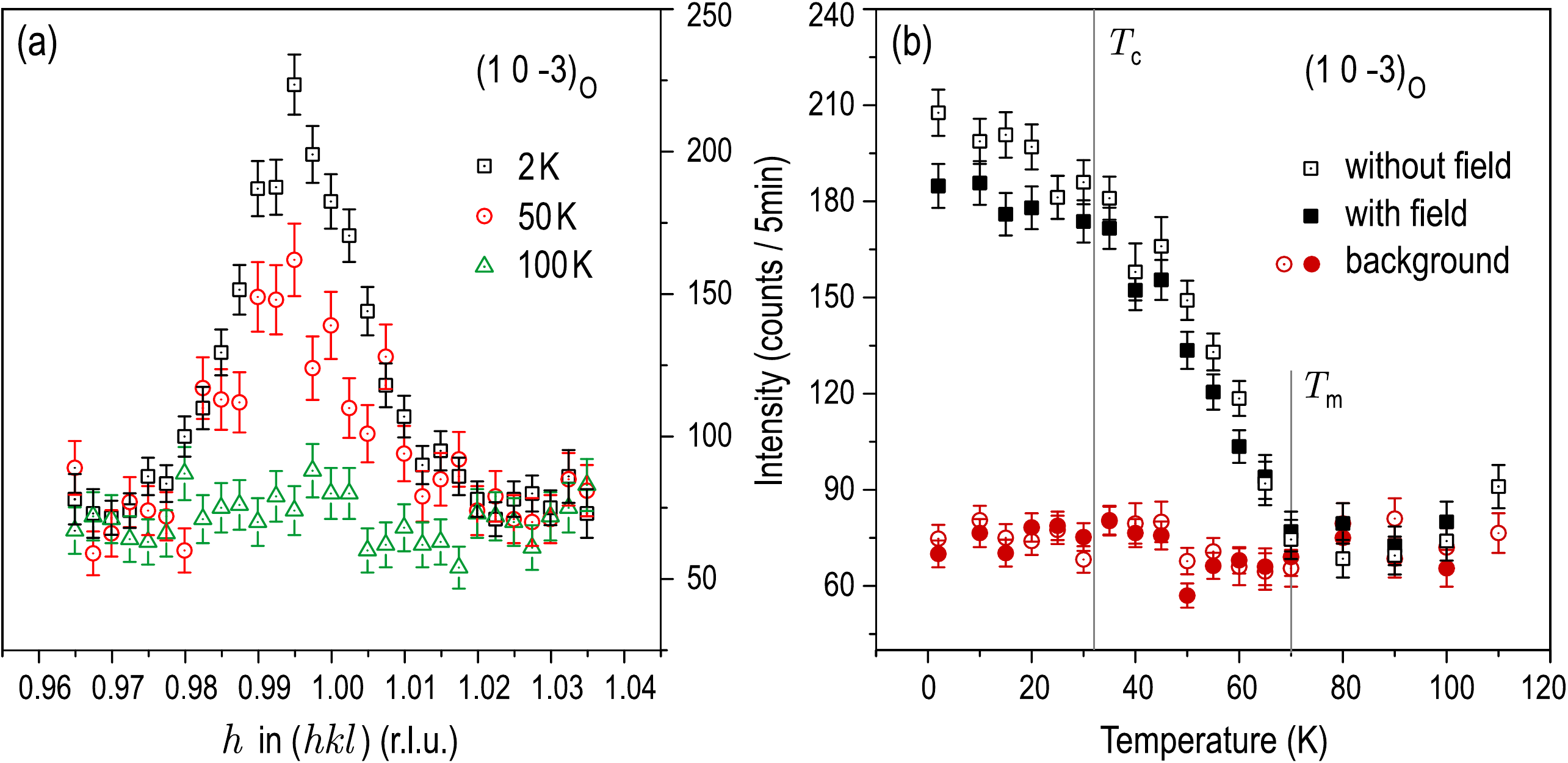}
\caption{Elastic neutron scattering data measured in the vicinity of the (1\,0\,--3)$_{\rm O}$ magnetic Bragg peak (O stands for orthorhombic notation). (a)~Scans along \mbox{($h$\,0\,--3)$_{\rm O}$} at three different temperatures. (b)~Temperature evolution of the magnetic intensity. The solid symbols are measured in the magnetic field of 13.5\,T applied parallel to the FeAs-layers.\vspace{-1.5em}}
\label{fig:neutron}
\end{figure}

Fig.\,\ref{fig:neutron} shows neutron scattering intensity measured in the vicinity of the (1\,0\,--3)$_{\rm O}$ magnetic Bragg peak \cite{Su08, Zhao08a} on a $\sim$\,30\,mg sample with in-plane and out-of plane mosaicities <\,1.5$^\circ$ and <\,2.5$^\circ$ respectively. The final neutron wave vector was set to $k_{\rm f}=1.55$\,${\text \AA}^{-1}$\hspace{-0.9em},\hspace{0.5em} and a Be-filter was used to extinguish contamination from higher-order neutrons. The sample was mounted with the orthorhombic $a$ and $c$ crystallographic directions in the scattering plane in a 15~T cryomagnet. Panel (a) shows \mbox{($h$\,0\,--3)$_{\rm O}$} scans at three different temperatures. While within the error bar there is no intensity at 100\,K, a clear magnetic peak starts to evolve at low temperatures. Panel (b) reveals the temperature evolution of the magnetic intensity, which lets us estimate the magnetic transition temperature $\TSDW\approx70$\,K.

From the width of the magnetic Bragg peaks, the lower estimate for the correlation length of the AF phase is $\zeta>100$\,\AA. This points to the first major difference in comparison with the underdoped cuprates. In \ybcosixfour, for example, the correlation length of the magnetic order is known to be much smaller, not exceeding 20\,\AA\ \cite{Hinkov08}.

Finally, we investigated the effect of the magnetic field $H = 13.5$\,T, applied perpendicular to the scattering plane and thus parallel
to the FeAs-layers. The magnetic intensity was suppressed by $\sim$\,10\%, as shown by solid symbols in Fig.\,2\,(b). This behavior, typical for an antiferromagnet, is again in contrast to the situation in cuprates, where the elastic intensity increases upon application of the magnetic field \cite{BfieldCuprates}, but is in-line with the notion of well-developed magnetic domains with commensurate AF order.

\begin{figure}[t]\vspace{0.3em}
\mbox{\hspace{-1ex}\includegraphics[width=1.02\columnwidth]{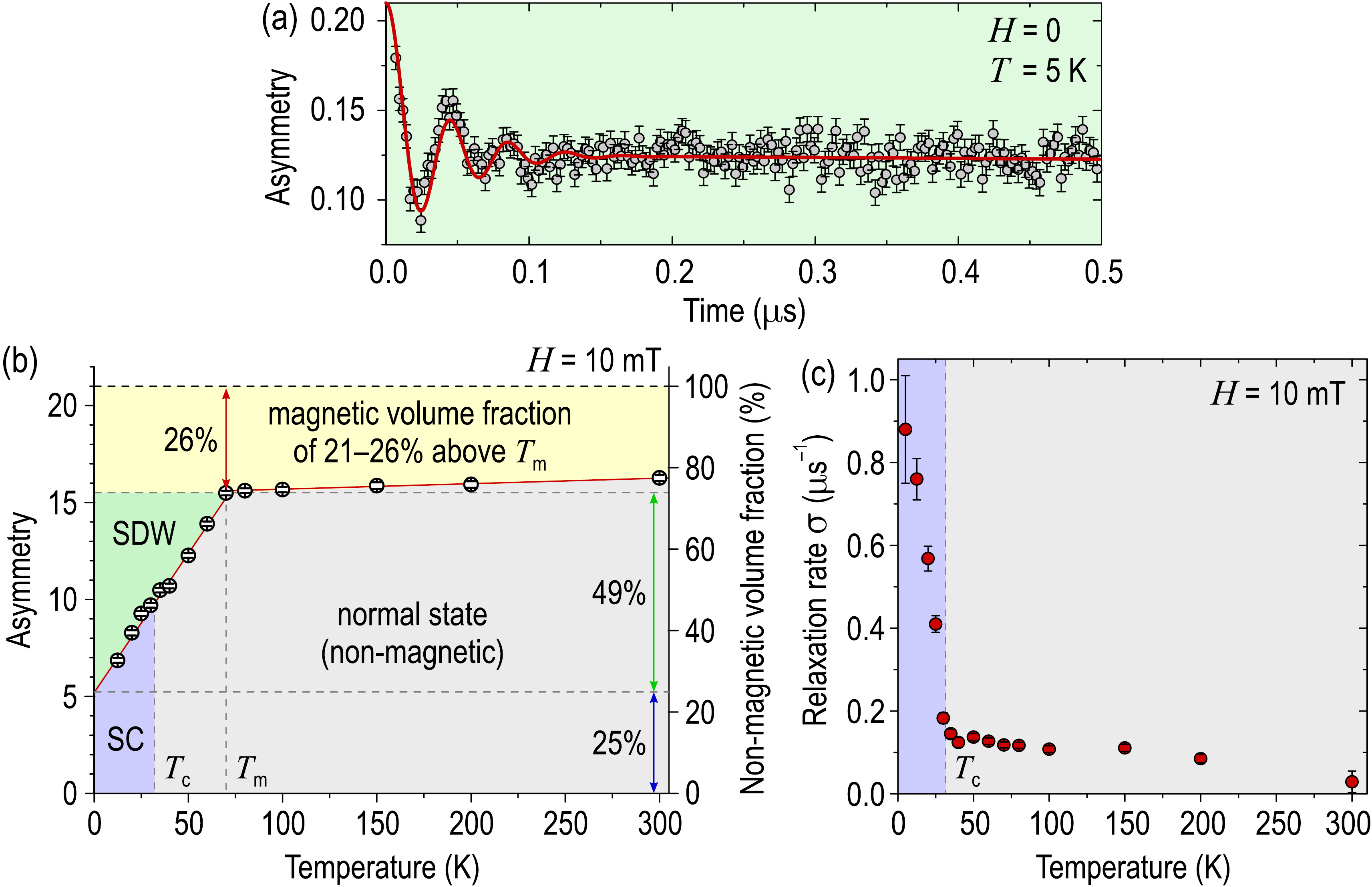}}\vspace{-0.2em}
\caption{\muSR\ data. (a)~Time dependence of the muon spin asymmetry in zero field. (b)~Temperature dependence of the asymmetry in a weak transverse field, showing coexistence of magnetic and non-magnetic phases. (c)~Temperature dependence of the relaxation rate in a transverse field.\vspace{-1.0em}}
\label{fig:muSR}
\end{figure}

To gain further insight into the nature of the magnetic ordering\,---\,in particular the magnitude of the ordered moment and the magnetic volume fraction\,---\,we performed zero-field (ZF) and transverse-field (TF) muon spin rotation (\muSR) measurements using $100$\% spin polarized muons, which in our setup corresponds to the muon spin asymmetry of $21$\,\% \cite{Brewer94}. The results of our \muSR\ measurements are illustrated by Fig.\,\ref{fig:muSR}. Panel (a) shows the time dependence of the asymmetry, which is a measure for the spin polarization of the muon ensemble. In principle, the oscillation frequency \zff\ is determined by the ordered Fe moment \mFe. Since the stopping position of the muon in the lattice is not known precisely, we resort to a comparison with \bfa, where \mFe\ was determined to be $0.4\mu_{\rm B}$ \cite{RotterPRB08}. The zero-field frequency for \bfa\ has been established to be $\zff=28$\,MHz. In comparison, for our sample $\zff=24.7(5)$\,MHz, so we estimate the ordered moment to be only slightly reduced to $\sim$\,0.35\,$\mu_{\rm B}$. This is remarkable, since simultaneously \TSDW\ is reduced by a factor of two from 140\,K to 70\,K.

By applying a weak field of $H=10$\,mT transverse to the original muon spin polarization, we can determine the non-magnetic volume fraction, in which the muons precess around $\mathbf{H}$ conserving the asymmetry, and the magnetically ordered fraction, in which a superposition of external and internal fields depolarizes the beam. Fig.\,\ref{fig:muSR}\,(b) shows the temperature dependence of the asymmetry in the transverse field. Surprisingly, already at 300\,K we observe a $\sim$\,21\% loss of asymmetry that might be an indication of a disordered magnetic phase. A straightforward explanation for it would be the presence of a magnetic impurity phase in our sample, such as Fe$_2$As ($T_{\rm N}=353$\,K), but such explanation can be ruled out, since XRPD performed on several pieces of samples from the same batch, ground into powder, indicated no presence of parasitic phases, as discussed above. Additionally, angle-resolved photoemission spectroscopy (ARPES) indicates the presence of some kind of density-wave-like order above \TSDW\ in the same samples, which is weakly temperature-dependent \cite{ZabolotnyyInosov08}. Assuming its magnetic character, it could be speculated that such ``hidden'' order is possibly responsible for the high-temperature loss of asymmetry observed by \muSR, which also decreases slightly with temperature above \TSDW.

Below $\sim$\,70\,K\,---\,the onset temperature of the magnetic intensity at the (1\,0\,--3)$_{\rm O}$ position\,---\,the asymmetry further decreases gradually from 15.5\% at \TSDW\ to 5.2\% at $T\rightarrow0$, indicating that the volume fraction of the SDW state is $\sim$\,49\% in the low-temperature limit. The remaining 25\% of the volume phase which remains non-magnetic at low temperature can be associated with the SC phase. For comparison, in nearly optimally doped BKFA with $x=0.5$, the low-temperature non-magnetic volume fraction constitutes almost 50\% \cite{Goko08}, in-line with the increased $T_{\rm c}=37$\,K. The SC volume fraction in our samples was also independently estimated from ARPES \cite{EvtushinskyInosov08}, which yielded $23\pm3$\% in agreement with our \muSR\ result.

Note that the decrease in asymmetry below \TSDW\ is gradual, indicating that we are dealing with a crossover rather than a sharp phase transition. This agrees with the absence of any appreciable anomalies at \TSDW\ in susceptibility and resistivity measurements.

Finally, we have measured the \muSR\ relaxation rate in the same transverse field. The weak magnetic field penetrates the sample through the AF islands, creating inhomogeneous field distribution within the SC phase, which results in rapid increase of muon depolarization below $T_{\rm c}$, as seen in Fig.\,\ref{fig:muSR}\,(c). Thus, the AF islands act as pre-formed vortex cores, precluding the formation of an ordered vortex lattice. At $T\rightarrow 0$ the relaxation rate, which in a homogeneous superconductor is expected to be proportional to the superfluid density according to the Uemura relation \cite{UemuraLuke89}, extrapolates to $\sigma=0.9\pm0.1$\,$\mu$s$^{-1}$. Surprisingly, this value follows the Uemura relation reasonably well, despite the phase separation. We note that our value of $\sigma$ is higher than that reported for the $x=0.45$ sample in Ref.~\onlinecite{Aczel08}, but still somewhat lower than in the optimally-doped $x=0.5$ sample from Ref.\,\onlinecite{Goko08}.

\begin{figure}[t]
\includegraphics[width=\columnwidth]{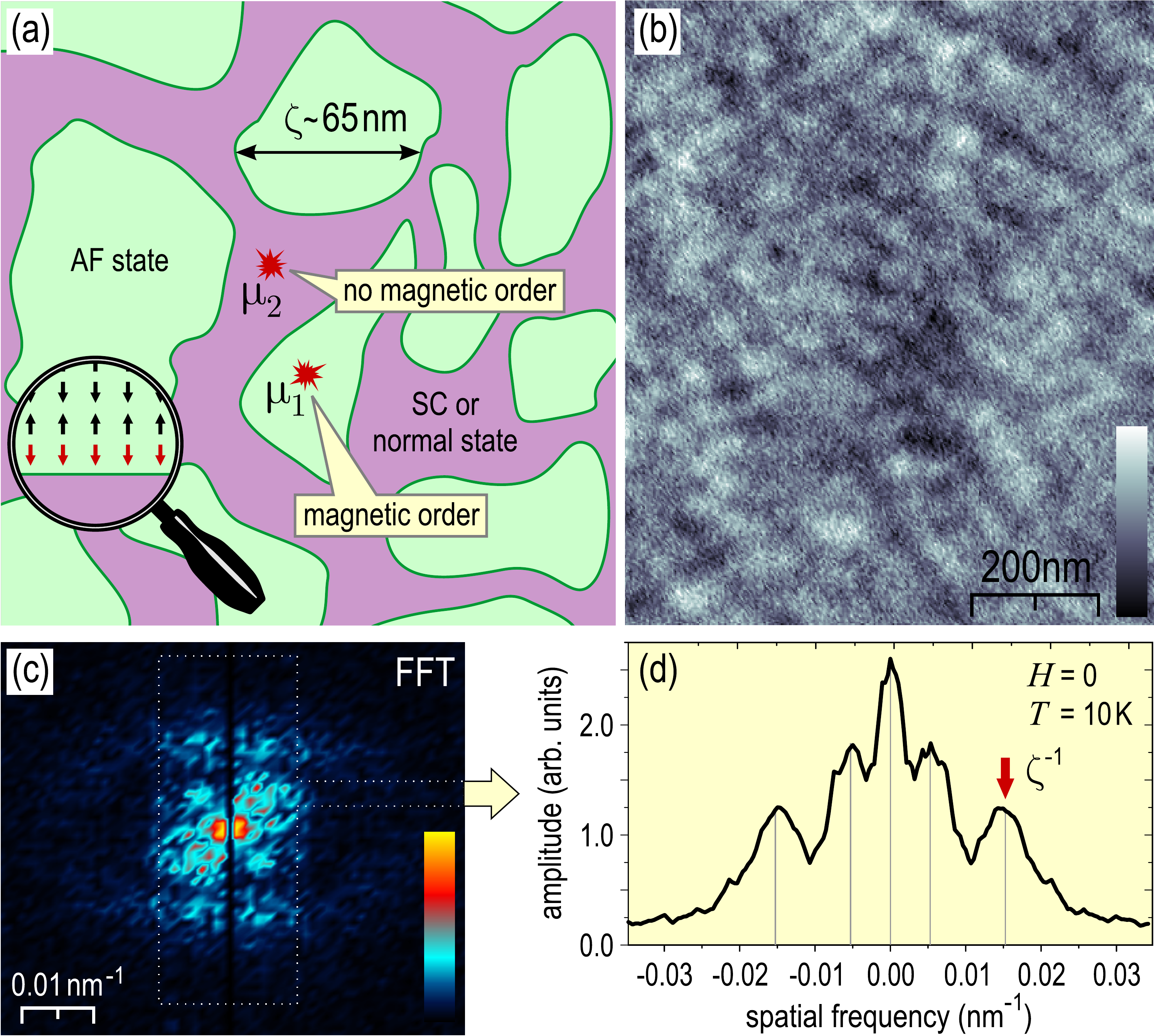}
\caption{(a)~Cartoonish representation of the phase-separated coexistence of AF and SC/normal states. (b)~MFM image measured at 10\,K in the absence of external field, revealing weak magnetic contrast on the lateral scale of $\sim$\,65\,nm, as can be estimated from the Fourier-transformed image in panel (c). Panel (d) shows the corresponding spatial frequency profile integrated within the dotted rectangle. The arrow marks the highest-frequency peak in the spectrum, responsible for the 65\,nm modulations.\vspace{-1.3em}}
\label{fig:MFM}
\end{figure}

At this point, we can already conclude that our sample simultaneously exhibits bulk SC with a sharp transition temperature of 32\,K and SDW order with a large correlation length $>100$\,\AA, which are spatially separated and change their volume ratio as a function of temperature. This resembles the situation in underdoped cuprates, where SC coexists with a short-range AF-correlated magnetic state with albeit strongly reduced ordered magnetic moment \cite{Niedermayer98}. There, however, the magnetic volume fraction seen by \muSR\ is nearly 100\% \cite{Niedermayer98}, indicating a more homogeneous coexistence of the two phases. On the other hand, scanning tunneling spectroscopy measurements provide numerous evidence for nano-scale inhomogeneities in the electronic density of states \cite{SeparationCuprates}. In contrast to the cuprates, in BKFA we rather observe a mesoscopic phase-separated coexistence \cite{Goko08}, as we schematically illustrate in Fig.\,\ref{fig:MFM}\,(a), with an ordered moment which is hardly suppressed as compared to the parent compound exhibiting long-range SDW order.

To get a better understanding of the real-space distribution of the magnetically ordered domains, we performed zero-field magnetic force microscopy (MFM) measurements in the SC state on a cleaved surface of a BKFA sample with somewhat reduced $T_{\rm c}$ of 26\,K using an \textit{Omicron Cryogenic SFM} scanning force microscope supplied with a commercial \textit{Nanoworld MFMR} magnetic tip possessing a force constant of $\sim$\,2.8\,N/m and a resonance frequency of 72\,kHz. Magnetic contrast was imaged with the lateral resolution <\,50\,nm by measuring the frequency shift at a scan height of 10\,nm above the sample surface. As shown in Fig.\,\ref{fig:MFM}\,(b), weak static magnetic contrast is clearly seen below $T_{\rm m}$, which would not be expected for a magnetically homogeneous sample. Successive scanning of the same area of the sample confirmed the reproducibility of the magnetic contrast at temperatures below \TSDW. The contrast is weakened above $T_{\rm m}$, though does not disappear completely. We associate this contrast with AF domain boundaries like those sketched in the inset in Fig.\,\ref{fig:MFM}\,(a), as the stray field produced by uncompensated magnetic moments at such a boundary is likely to result in a magnetic contrast detectable by MFM. To estimate the characteristic spatial scale of the observed inhomogeneities, we performed a Fourier transform of the MFM signal [see Fig.\,\ref{fig:MFM}\,(c) and (d)]. The highest-frequency peak in the spectrum corresponds to the characteristic scale of the inhomogeneities of the order of $\zeta=65\pm10$\,nm. A peak corresponding to larger-scale modulations can also be seen in some of the spectra.

To summarize, we have observed mesoscopic phase-separated coexistence of magnetically ordered and non-magnetic states on the lateral scale of $\sim$\,65\,nm in a slightly underdoped iron pnictide superconductor, as estimated from MFM imaging in agreement with the \muSR\ measurements. Though phase separation is clearly an intrinsic property of the studied material, it still remains to be investigated to which extent it correlates with the distribution of the dopants. The scale of the electronic inhomogeneities is reminiscent of the situation in superoxygenated La$_2$CuO$_{4+\delta}$, where phase separation happens on similar scales of 30\,--\,300\,nm \cite{SeparationMacro}, but is markedly different from the quasi-homogeneous nano-scale mixture of electronic phases in most other cuprates \cite{SeparationCuprates}.

\textit{Acknowledgements.} The experimental work was performed at the Swiss spallation source SINQ (\textit{RITA-II} spectrometer), the Swiss Muon Source, both at the Paul Scherrer Institut (PSI), Villigen, Switzerland, and the X16C beamline at the National Synchrotron Light Source, Brookhaven National Laboratory, USA. MFM measurements were funded by the European Community under the 6th Framework Programme Contract No.\,516858: HIPERCHEM. Image processing was done using the WSxM 4.0 software \cite{Horcas07}. We acknowledge financial support from DFG in the consortium FOR538, as well as from FCI and BMBF. We also thank P.~W.~Stephens for the assistance with XRPD measurements and A.~Leineweber from MPI-MF for helpful discussions.

\end{document}